\documentclass[twocolumn,groupedaddress,showpacs,nofootinbib,aps,pra]{revtex4-1}

\usepackage[T1]{fontenc}
\usepackage[cp1250]{inputenc}
\usepackage{graphicx}
\usepackage{dcolumn}
\usepackage{bm}
\usepackage{ulem}
\usepackage{amsmath}

\newcommand{\vect}[1]{\boldsymbol{#1}}
\newcommand*\colvec[3][]{
    \begin{pmatrix}\ifx\relax#1\relax\else#1\\\fi#2\\#3\end{pmatrix}
}

\begin{document}
\title{Dichroic atomic vapor laser lock with multi-gigahertz stabilization range}

\author{S. Pustelny}
 \email{pustelny@uj.edu.pl}
 \affiliation{Institute of Physics, Jagiellonian University, {\L}ojasiewicza 11, 30-348 Krak\'ow, Poland}
 \affiliation{Department of Physics, University of California at Berkeley, Berkeley, 94720-7300, USA}

\author{V. Schultze}
 \affiliation{Leibniz Institute of Photonic Technology, Albert-Einstein-Str. 9, D-07745 Jena, Germany}

\author{T. Scholtes}
 \affiliation{Leibniz Institute of Photonic Technology, Albert-Einstein-Str. 9, D-07745 Jena, Germany}

\author{D. Budker}
 \affiliation{Helmholtz-Institut Mainz, Johannes Gutenberg Universit\"at Mainz, 55128 Mainz, Germany}
 \affiliation{Department of Physics, University of California, Berkeley, CA 94720-7300, USA}

\begin{abstract}
A dichroic atomic vapor laser lock (DAVLL) system exploiting buffer-gas-filled millimeter-scale vapor cells is presented. This system offers similar stability as achievable with conventional DAVLL system using bulk vapor cells, but has several important advantages. In addition to its compactness, it may provide continuous stabilization in a multi-gigahertz range around the optical transition. This range may be controlled either by changing the temperature of the vapor or by application of a buffer gas under an appropriate pressure. In particular, we experimentally demonstrate the ability of the system to lock the laser frequency between two hyperfine components of the $^{85}$Rb ground state or as far as 16 GHz away from the closest optical transition.
\end{abstract}


\maketitle

\section{Introduction}

Most of modern atomic, molecular, and optical physics experiments require precise control over optical properties of light. For example, long-term stability of light frequency is needed in experiments with optical cooling and trapping, investigation of nonlinear (coherent) optical effects, optical metrology (e.g., atomic magnetometry), or searches for physics beyond the Standard Model. A popular technique of laser-frequency stabilization exploits magnetically induced circular anisotropy of a medium \cite{Cheron1993Laser}. In this technique, a longitudinal magnetic field, i.e., field along the quantization axis, shifts energies of Zeeman sublevels lifting their degeneracy. This results in a frequency-dependent difference in absorption and dispersion of the two orthogonal circular polarization ($\sigma^{\pm}$) components of light\footnote{The quantization axis is oriented along the light-propagation direction.}. Thus, detection of light ellipticity or polarization rotation provides an error signal, which enables light-frequency stabilization. This approach has been used in various incarnations of dichroic atomic vapor laser locks (DAVLLs) employed for stabilization of laser-light frequency to optical transitions in helium \cite{Cheron1993Laser}, alkalis (rubidium \cite{Corwin1998Frequency,Yashchuk2000Laser,Wasik2002Laser,Petelski2003Doppler}, cesium \cite{Clifford2000Stabilization,Beverini2003Frequency,Overstreet2004Zeeman}, potassium \cite{Pichler2012Simple}), strontium \cite{Javaux2010Modulation}, mercury \cite{Yin2012Observation,Liu2013Optimization}, ytterbium \cite{Kim2003Frequency}, and barium \cite{Hosegawa2009Laser}.

One of the most important advantages of the DAVLL system is its technical simplicity. A complete DAVLL set consists of a vapor cell, magnets or a solenoid, a pair of wave plates (quarter- and half-wave plate), a polarizer with two photodiodes (polarimeter), and a proportional-integral-derivative (PID) controller to process electrical signals. This limited number of components enables miniaturization of the system. In fact, a DAVLL with a volume of several cm$^3$ was recently demonstrated \cite{Lee2011Small}. The key element of the system was a microfabricated vapor cell (microcell) with a volume smaller than $1$~mm$^3$ \cite{Liew2003Microfabricated}. As shown in Ref.~\cite{Lee2011Small}, the performance of such a system is comparable to that of a traditional DAVLL, which, among the others, opens the possibility of its incorporation into laser heads to enhance the laser performance or construction of portable atomic magnetometers \cite{Schwindt2007Chip}

Another important feature of the DAVLL system is its ability to lock the laser in a relatively broad frequency range (e.g., $\sim600$~MHz in systems exploiting a buffer-gas-free room-temperature alkali vapor). While this is significantly more than obtainable with other techniques (e.g., those based on saturated-absorption spectroscopy or those referencing the laser to an optical cavity), it may not be enough for some applications. For example, pump-probe spectroscopy requires detuning of a probe laser several linewidths from the investigated transition. In such a case, off-resonance locking may be achieved by superimposing two light beams, one to be stabilized and other that is stabilized to a particular transition, and measuring/locking their beating frequency \cite{Schunemann1999Simple}. Alternatively, laser stabilization may be realized by seeding a slave laser with a frequency-stabilized laser, whose amplitude or phase is modulated with an electro-optical modulator (EOM) \cite{Szymaniec1997Incjection}. The drawback of these solutions, however, is their complexity, as they require two lasers and EOM, and detection of beating (error) signals at a frequencies often exceeding 1~GHz.

In general, off-resonance locking may be also achieved with the DAVLL system. For such a purpose, however, the DAVLL locking range needs to be shifted or expanded to include the frequency of interest.

From a practical point of view, the locking range in DAVLL systems is limited by the width of the optical transition used for stabilization. Therefore, a straightforward approach to off-resonance stabilization would exploit broadening of the absorption line in the locking system. This may be achieved either by heating the vapor cell (increasing absorption) or by introduction of an additional gas into the cell (inducing pressure broadening). The first approach was recently demonstrated in the context of the Faraday spectroscopy, where laser locking 6-14~GHz away from the optical transition was demonstrated \cite{Marchant2011Off}. While this solution enables stabilization of laser frequency far from the optical transition, it does not provide an opportunity to lock the laser closer to the transition, where absorption is so strong that no light is transmitted through the cell\footnote{In order to lock the laser closer to the optical transition the temperature of the cell needs to be reduced.}.

In this paper, we investigate light-frequency stabilization employing the combination of the approaches based on increasing absorption and inducing pressure broadening. To do that we build a DAVLL system with microcells filled with rubidium vapor and molecular nitrogen as a buffer gas. This allows us to theoretically estimate the ability to lock the laser in a 40-GHz continuous frequency range, and experimentally demonstrate locking the laser in a continuous  30-GHz range extending from 16~GHz below the center of the $F=1\rightarrow F'=2$ transition to 8~GHz above the center of the $F=2\rightarrow F'=1$ transition of the $^{87}$Rb $D_1$ line. In particular, locking the laser in the middle between two Doppler-broadened transitions of $^{85}$Rb ($F=3\rightarrow F'$ and $F=2\rightarrow F'$), as well as, 16 GHz away from the optical transition is presented. In all the cases, compensation of $\approx$1-GHz drifts or instant frequency jumps with a residual instability less than 5~MHz is demonstrated. The performance of the system is investigated as a function of temperature of the vapor and pressure of the buffer gas. To confirm the capabilities of the technique, we support experimental results with numerical simulations based on the density-matrix formalism.

\section{Experimental setup}

\begin{figure}
    \includegraphics[width=\columnwidth]{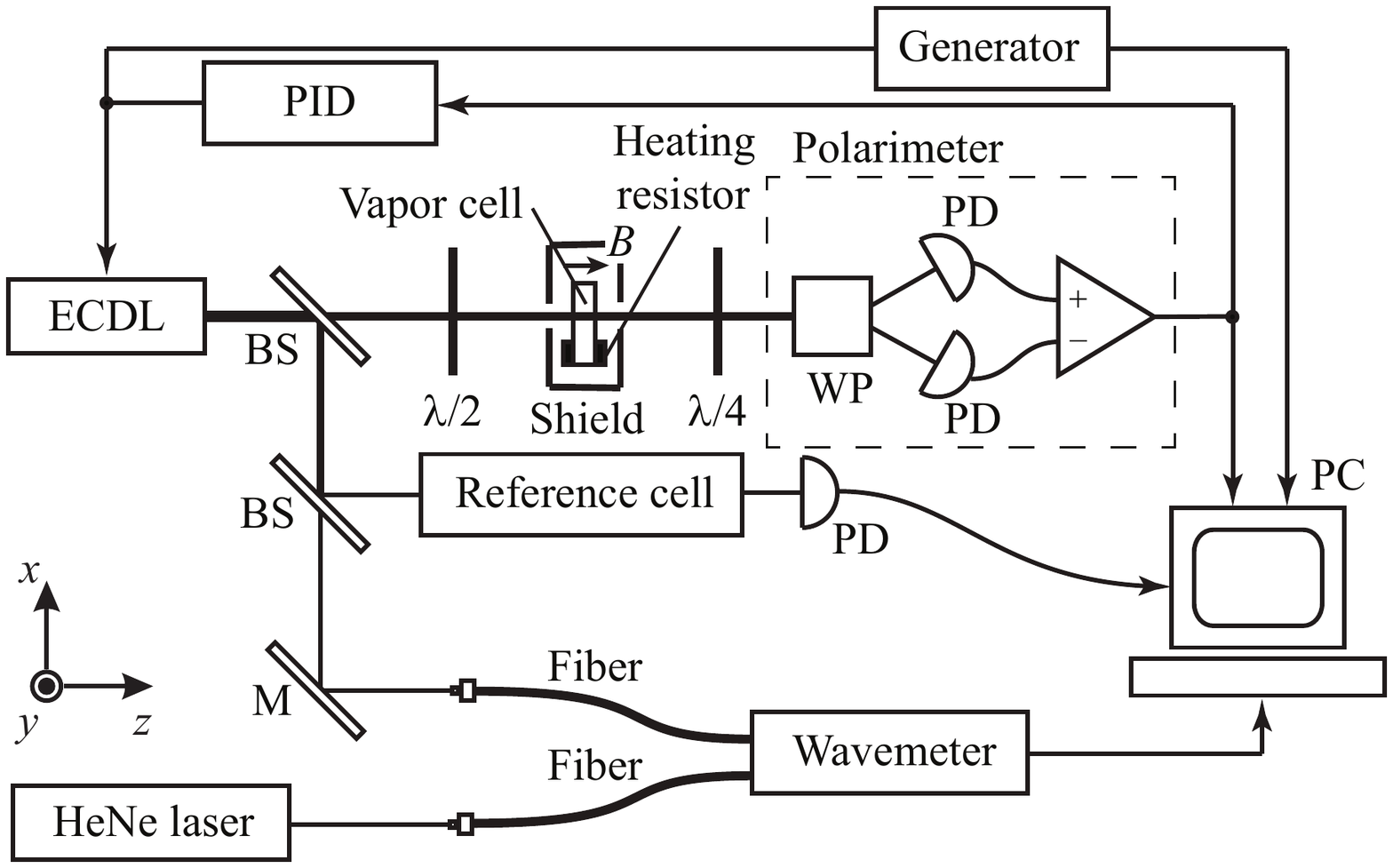}
    \caption{Schematic of the experimental setup used to characterize the DAVLL system. ECDL stands for the extended-cavity diode laser, HeNe laser is a frequency-stabilized HeNe laser operating as the reference for the wavemeter, BS stands for a beam-splitter, M is a mirror, WP is a Wollaston prism, PD is a photodiode, and $\lambda/4$ and $\lambda/2$ are quarter- and half-wave plates, respectively.\label{fig:Setup}}
\end{figure}

The experimental arrangement used for laser-frequency locking and characterization of the DAVLL performance is shown in Fig.~\ref{fig:Setup}.
The DAVLL system was essentially identical to the one used in Ref.~\cite{Lee2011Small}. The only difference was in a microfabricated rubidium (natural abundance) vapor cell ($1\times 2\times 3$~mm$^3$), manufactured at IPHT Jena as described in Ref.~\cite{Woetzel2011Microfabricated} for the case of cesium. In addition to rubidium, the cells used in this work contained molecular nitrogen, as a buffer gas, at two pressures: 46~mbar and 230~mbar \footnote{Buffer-gas pressure was determined based on Lorentzian fits to the $D_2$ absorption spectra, using the relation between frequency shift and buffer-gas pressure of -4.52 MHz/mbar \cite{Romalis1997Pressure}.}. The cells were heated up to 125$^\circ$C with a set of resistors. The vapors were subject to a 200-G longitudinal magnetic field (with less then 20~G inhomogeneity within the cell volume \cite{Lee2011Small}) generated with a set of toroidal permanent magnets. The cell's heater-magnet system was enclosed in a single-layer $\mu$-metal cylindrical magnetic shield with two lids of a total length of 16~mm and a diameter of 9~mm.

For a thorough characterization of the microDAVLL system, additional measurements were performed in a ``traditional'' DAVLL system, employing a bulk cylindrical evacuated vapor cell of a length of 10 mm and 10~mm in diameter \cite{Yashchuk2000Laser}. The bulk system was operated under the conditions providing comparable properties to those of the microDAVLL, particularly, optical depth, magnetic field, and light intensities.

To control the locking point, a pair of wave plates was used in the systems: a half-wave plate was mounted in front of the cell and a quarter-wave plate behind the cell. This provides the control over the stabilization point via adjustment of the DAVLL-signal zero crossings. Particularly, the wave plates allow for switching between the dichroism detection, enabling locking close to the center of the transition, and the Faraday-rotation detection (birefringence), when laser light can be locked far from the transition. Moreover, the two wave plates can be used to eliminate thermal drifts of the locking point \cite{Kostinski2011Temperature}.

After the shield, the light was detected with a balanced polarimeter, consisting of a Wollaston prism and two photodiodes. The polarimeter output signal (a photodiode-difference signal) was amplified and fed into a PID system (SRS SIM960) and next into the modulation input of the laser controller (External Cavity Diode Laser from VitaWave). For the characterization of the system, the PID signal was combined with an external-generator signal. It allows investigation of the system's ability to compensate for drifts or instant changes (square-wave modulation) of the laser frequency. The laser frequency was monitored with a wavemeter (HighFinesse WS/7), which according to the specification, offered a spectral accuracy of 200~MHz. In our measurements, higher resolution of the measurements was achieved by referencing the wavemeter to a frequency-stabilized He-Ne laser (Spectra Physics 117A with a residual instability less than 5~MHz over 1 hour). In such a way, a residual instability of the wavemeter of about 1~MHz within half an hour was demonstrated.

\section{Simulation of the DAVLL signal\label{sec:Simulations}}

In this work, the experimental investigations are supported with theoretical modeling of the system. The modeling is based on the density-matrix calculations exploiting a full experimental system with two hyperfine ground states and two hyperfine excited states of two rubidium isotopes ($^{85}$Rb: $F=2,3$ and $F'=2,3$ and $^{87}$Rb: $F=1,2$ and $F'=1,2$) with a complete set of magnetic sublevels, interacting with linearly polarized light. The atoms are subject to a longitudinal magnetic field splitting Zeeman sublevels and mixing those of the same $m$ but different $F$, where $m$ is the magnetic quantum number. The interaction with light and magnetic field results in optical anisotropy (circular birefringence and dichroism) of the medium; polarization rotation and ellipticity change of light propagating through the medium is observed.

The Hamiltonian of the system is given by
\begin{equation}
	\hat{H}=\hat{H}_0-\hat{\vect{d}}\cdot\vect{E}-\hat{\vect{\mu}}\cdot\vect{B},
\end{equation}
where $\hat{H}_0=\sum_{Fm}\omega_{F}|Fm\rangle\langle Fm|$ is the Hamiltonian of an unperturbed system with $\omega_F$ being the frequency of a hyperfine level with the total angular momentum $F$, $\vect{E}$ and $\vect{B}$ are electric and magnetic fields, respectively and $\hat{\vect{d}}$ and $	\hat{\vect{\mu}}$ are respective electric- and magnetic-dipole-moment operators (we choose a natural unit system with $c=\hbar=1$). The evolution of the density matrix is governed by the Liouville equation
\begin{equation}
	\dot{\hat{\rho}}=-i\left[\hat{H},\hat{\rho}\right]-\hat{\Gamma}({\hat{\rho}})+\hat{\Lambda}(\hat{\rho}),
\end{equation}
where $\hat{\Gamma}(\hat{\rho})$ is the relaxation operator and $\hat{\Lambda}(\hat{\rho})$ is the repopulation operator. There are two sources of relaxation/repopulation in the system under consideration. First is uniform relaxation that tends to repopulate atoms toward thermal equilibrium (equally populated ground-state magnetic sublevels of a given isotopes and 27/73 division of the population between $^{87}$Rb and $^{85}$Rb determined by the isotope abundances). The process is induced mainly by collisions with the walls of the cell. The second relaxation/repopulation process is associated with optical pumping and spontanous emission, which drives the system toward dynamic equilibrium incorporating both processes.

By introducing the macroscopic polarization $\vect{P}$ and relating it to the electric field $\vect{E}$ and the density matrix $\rho$, $\vect{P}=n_{at}\textrm{Tr}\left(\hat\rho \hat{\vect{d}}\right)$, where $n_{at}$ is the rubidium number density, one can derive the equations for the change in electric field of light $E_0$, light polarization rotation $\varphi$, and ellipticity change $\varepsilon$ per unit distance \cite{Auzinsh2010Optically}
\begin{widetext}
	\begin{equation}
		\begin{aligned}
		\frac{dE_0}{dl}(\omega)&=& -\pi\omega n_{at}\textrm{Im}\left[\sum_{FmF'}\|d_{FF'}\|\left(\langle Fm11|F'\ m\!+\!1\rangle 					\rho_{Fm,F'm+1}+  \langle Fm1-\!1|F'\ m\!-\!1\rangle \rho_{Fm,F'm-1} \right)\right], \\
		\frac{d\varphi}{dl}(\omega)&=& -\frac{\pi\omega n_{at}}{E_0}\textrm{Re}\left[\sum_{FmF'}\|d_{FF'}\|\left(\langle Fm11|F'\ m\!+\!1\rangle 					\rho_{Fm,F'm+1}+  \langle Fm1-\!1|F'\ m\!-\!1\rangle \rho_{Fm,F'm-1} \right)\right], \\
		\frac{d\varepsilon}{dl}(\omega)&=& \frac{\pi\omega n_{at}}{E_0}\textrm{Im}\left[\sum_{FmF'}\|d_{FF'}\|\left(\langle Fm11|F'\ m\!+\!1\rangle 						\rho_{Fm,F'm+1}+  \langle Fm1-\!1|F'\ m\!-\!1\rangle \rho_{Fm,F'm-1} \right)\right],
		\end{aligned}
		\label{eq:RotationEllipticity}
	\end{equation}
\end{widetext}
where $\|d_{FF'}\|$ is the reduced electric dipole moment between the hyperfine ground state ($F$) and hyperfine excited state ($F'$), $\lambda$ is the wavelength of light, and $\langle ... | ...\rangle$ is the Clebsch-Gordan coefficient. Summation in Eqs.~(\ref{eq:RotationEllipticity}) runs over all ground-state magnetic sublevels.

It should be noted that Eqs.~(\ref{eq:RotationEllipticity}) are derived for $x$-polarized light. This allows one to present them in a relatively compact form. In general, however, light polarization is arbitrarily oriented in the $xy$-plane, which is determined by the orientation of the half-wave plate situated in front of the cell. Nonetheless, since there is no preferred direction in the plane, the means of light-atom interaction (i.e., light absorption, polarization rotation, ellipticity change) are identical for all linear polarization in the $xy$-plane. Thus, in the calculations, the effect of arbitrary orientation of light polarization is incorporated by rotating the polarization axis of light departing the atomic medium by the angle $2\theta_{\lambda/2}$ rather then rotating polarization of the incident light\footnote{An explicit form for polarization absorption, rotation, and change in ellipticity of arbitrarily polarized light can be found in Ref.~\cite{Auzinsh2010Optically}.}.

In alkali-vapor cells, there are two contributions to the linewidths of optical transitions, which are important for determination of the DAVLL signal. The first contribution stems from the finite lifetime of an atom in the excited state. In the buffer-gas cells, the lifetime is predominantly determined by the alkali atoms collisions with the buffer-gas atoms/molecules, while in the buffer-gas-free cells it is determined by spontaneous emission. The buffer-gas collisions introduce additional factor to the excited-state relaxation, which enters Eqs.~(\ref{eq:RotationEllipticity}) via density-matrix element $\rho_{Fm,F'm'}$ but also through the modifications of reduced electric dipole moment $\|d_{FF'}\|$. Nonetheless, as the collisions affect all atoms identically, i.e., each atom reveals the same excited-state relaxation rate $\gamma_e$, the collisions lead to homogenous broadening of the transitions. The second contribution to the linewidth stems from a thermal motion of atoms; the Doppler shift that is associated with the atomic/molecular motion results in an inhomogeneous broadening of the transitions. As the shift is specific for atoms moving with specific velocities, this effect is not present in Eqs.~(\ref{eq:RotationEllipticity}), it can be, however, included by convolving them with a Doppler broadened line
\begin{equation}
	\begin{aligned}
		\left<\frac{dE_0}{dl}(\omega)\right>_{v}&=&\frac{1}{\sqrt{\pi}\Delta_D}\int_{-\infty}^{\infty}\frac{dE_0}{dl}\left(\omega-\omega'\right)e^{-\omega'^2/\Delta_D^2}d\omega', \\
		\left<\frac{d\varphi}{dl}(\omega)\right>_{v}&=&\frac{1}{\sqrt{\pi}\Delta_D}\int_{-\infty}^{\infty}\frac{d\varphi}{dl}\left(\omega-\omega'\right)e^{-\omega'^2/\Delta_D^2}d\omega', \\
		\left<\frac{d\varepsilon}{dl}(\omega)\right>_{v}&=&\frac{1}{\sqrt{\pi}\Delta_D}\int_{-\infty}^{\infty}\frac{d\varepsilon}			{dl}(\omega-\omega')e^{-\omega'^2/\Delta_D^2}d\omega',
	\end{aligned}
	\label{eq:RotationEllipticityAveraged}
\end{equation}
where $\Delta_D$ is the Doppler width of all transitions\footnote{Although there are difference between frequencies of specific transitions, the difference are small compared to the transitions frequencies. Hence, herein, we assumed identical Doppler broadening of all optical transitions.}.

Finally, to calculate the polarization rotation induced by the medium, Eqs.~(\ref{eq:RotationEllipticityAveraged}) need to be integrated over the transition path through the medium. The result of this calculation provides parameters that enable one to determine the DAVLL signal.

To calculate a signal at the polarimeter (the DAVLL signal), we used the Jones formalism. Within the formalism, the signal at the balanced polarimeter is given by
\begin{equation}
	S=A\left(|S_x|^2-|S_y|^2\right),
	\label{eq:Signal}
\end{equation}
where $A$ is the factor depending on light intensity and gain in the detection system and $S_x$ and $S_y$ are the two elements of the Jones vector
\begin{widetext}
	\begin{equation}
		\colvec{S_x(\omega)}{S_y(\omega)}=M_\frac{\lambda}{4}(\theta_{\lambda/4})\colvec{-i\cos(\varphi_l(\omega)+2\theta_{\lambda/2})\sin\varepsilon_l(\omega)-\cos\varepsilon_l(\omega)\sin(\varphi_l(\omega)+2\theta_{\lambda/2})}{\cos(\varphi_l(\omega)+2\theta_{\lambda/2})\cos\varepsilon_l(\omega)-i\sin\varepsilon_l(\omega)\sin(\varphi_l(\omega)+2\theta_{\lambda/2})}e^{-\kappa_l(\omega)},
	\label{eq:JonesVector}
	\end{equation}
\end{widetext}
where $\kappa_l(\omega)$, $\varphi_l(\omega)$, and $\varepsilon_l(\omega)$ are the light-frequency dependent absorption, polarization rotation, and ellipticity change given by Eqs.~(\ref{eq:RotationEllipticityAveraged}) integrated along the light path (e.g., $\varepsilon_l(\omega)=\int_0^l\langle dE_0/dl'(\omega)\rangle_vdl'$) and $M_{\frac{\lambda}{4}}(\theta_{\lambda/4})$ is the matrix describing light polarization transformation induced by a quarter-wave plate oriented at the angle $\theta_{\lambda/4}$ with respect to the $x$ axis
\begin{equation}
	M_{\frac{\lambda}{4}}(\theta_{\lambda/4})=\left(\begin{matrix} \frac{1}{\sqrt{2}}(1+\cos 2\theta_{\lambda/4})&-\frac{i}{\sqrt{2}}\sin2\theta_{\lambda/4}\\
	-\frac{i}{\sqrt{2}}\sin2\theta_{\lambda/4}&\frac{1}{\sqrt{2}}(1-\cos 2\theta_{\lambda/4}) \end{matrix}\right).
	\label{eq:Quarterwaveplate}
\end{equation}
	
In this paper, Eqs.~(\ref{eq:Signal})-(\ref{eq:Quarterwaveplate}) are used to simulate the DAVLL signals.

\section{Results and analysis}

\begin{figure}
    \includegraphics[width=\columnwidth]{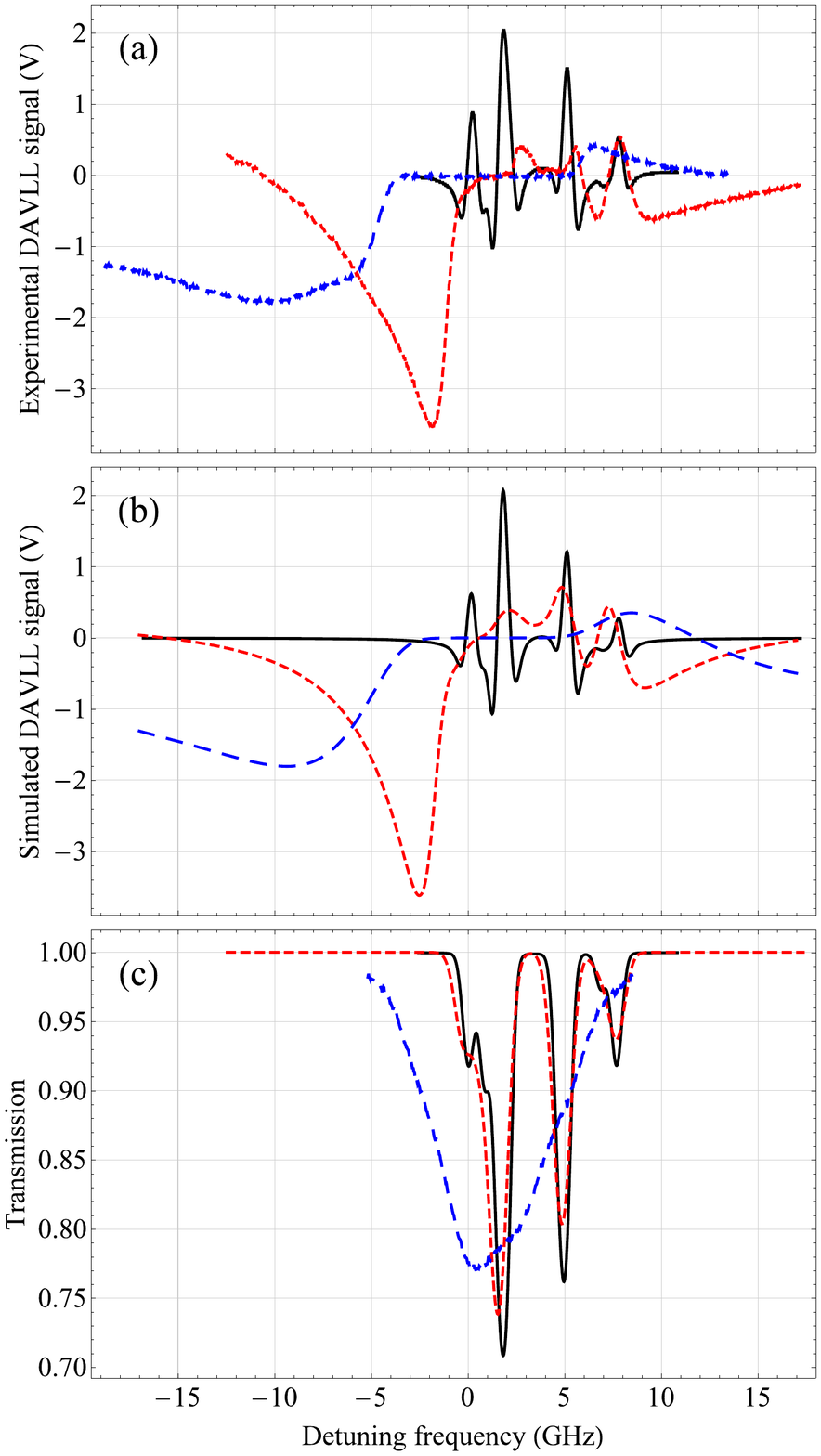}
    \caption{(a) Measured and (b) simulated DAVLL spectra recorded in bulk- (solid black) and two microcell (46 mbar of N$_2$ -- dotted red and 230~mbar of N$_2$ -- dashed blue) systems. The bulk-cell measurements were performed at $75^\circ$C, while the temperature of microcells were $\sim 100^\circ$C. The microDAVLL signals were measured with a 5-mW light beam. (c) The corresponding normalized absorption spectra in the cells (after subtraction of the linear background in the signal.) For better visibility of the spectrum, the two spectra recorded in buffer-gas microcells were measured at a slightly lower temperature (94$^\circ$C), so that light transmitted through the cell, close to the center of absorption lines, is still detectable (at 100$^\circ$C the medium is completely opaque for the tunings close to the center of the absorption lines).}
    \label{fig:ErrorSpectrumPressure}
\end{figure}

Figure~\ref{fig:ErrorSpectrumPressure} presents the DAVLL signals measured [Fig.~\ref{fig:ErrorSpectrumPressure}(a)] and simulated [Fig.~\ref{fig:ErrorSpectrumPressure}(b)] versus light frequency in three atomic vapor cells (two microcells and a bulk cell with no buffer gas)\footnote{The DAVLL signal in the bulk system was measured for the Faraday-spectroscopy arrangement (purely birefringence-determined signal), while the signals in the microcells are combination of dichroic and birefringent contributions.} together with the corresponding absorption spectra (measured at lower temperatures) [Fig.~\ref{fig:ErrorSpectrumPressure}(c)].

The presented data reveal differences among the three cells. Particularly, the number of zero crossings of the signals, i.e., the number of independent locking points, varies between the cells. While 10 zero crossings are observed in the buffer-gas-free cell, significantly fewer crossings are present in the microcell signals. This difference originates from a different ratio between linewidths of specific optical transitions and the level splittings. In particular, in the evacuated cell the Doppler width is on the order of 300~MHz, i.e., it has narrower than most of the hyperfine splittings. In turn, the transitions independently contribute to the signal. In the investigated buffer-gas-filled cells\footnote{The width of the rubidium $D_1$ transitions (full width at half maximum) due to the collisions with N$_2$ is calculated from a pressure broadening of $\sim 18$~MHz/mbar \cite{Romalis1997Pressure}.}, the pressure broadening overwhelms the transition linewidth (additionally causing deterioration of the transition amplitudes), which results in an overlap between neighboring transitions. Thereby, fewer zero-crossings are observed in the DAVLL signal.

Another source of difference between the signals arises due to different temperatures under which the DAVLL signals were detected. Specifically, the microcell signals were measured at temperatures, corresponding to stronger light absorption and hence larger amplitudes of the DAVLL signals are observed. Moreover, under such conditions, the medium is opaque for light tuned close to the strongest transitions (e.g., at the $^{85}$Rb transitions), thus no light and hence no DAVLL signal is observed for such tunings.

Other differences are observed at the wings of the signals. In particular, the negative-detuning wings of the signals measured in the buffer-gas cells extend much further than those of the evacuated cell. This stems from the higher temperature under which the cells are operated, i.e., the approach demonstrated, for example, in Ref.~\cite{Marchant2011Off}, but also from the pressure broadening and shift of the optical transitions. The impact of the buffer gas is also visible in the high-frequency wings of the DAVLL spectra. In particular, for this part of the spectrum the DAVLL signal measured in the 46-mbar and 230-mbar cells have opposite signs; in an optical arrangement used in Fig.~\ref{fig:ErrorSpectrumPressure}, the 230-mbar cell reveals only positive signal for all positive detunings (there are no zero crossings), while such crossings are observed in the two remaining cases.

Experimental results shown in Fig.~\ref{fig:ErrorSpectrumPressure}(a) are supported with theoretical simulations based on the density-matrix formalism described in Sec.~\ref{sec:Simulations} [Fig.~\ref{fig:ErrorSpectrumPressure}(b)]. The results  are in general agreement with the experimental data [Fig.~2(a)]. Particularly, the simulations show the change in the signal shape, including the number of zero crossings in different cells. The small differences in the shape of the experimental and theoretical signals is most likely a consequence of our experimental procedure. We scan laser frequency by changing laser-diode current. Thus the scan is accompanied by the changes of light intensity. It lead to appearance of a slope superimposed on the atomic signal or the additional dependence of the DAVLL signal on the light frequency [dependence of the factor $A$ in Eq.~(\ref{eq:Signal}) on the light frequency] and hence slight modification of the observed DAVLL spectra.

To further study the buffer-gas dependence of the DAVLL signal, we simulated the spectra for different buffer-gas pressures but otherwise same set of parameters (Fig.~\ref{fig:ConcentrationDependenceFull}).
\begin{figure}
	\includegraphics[width=\columnwidth]{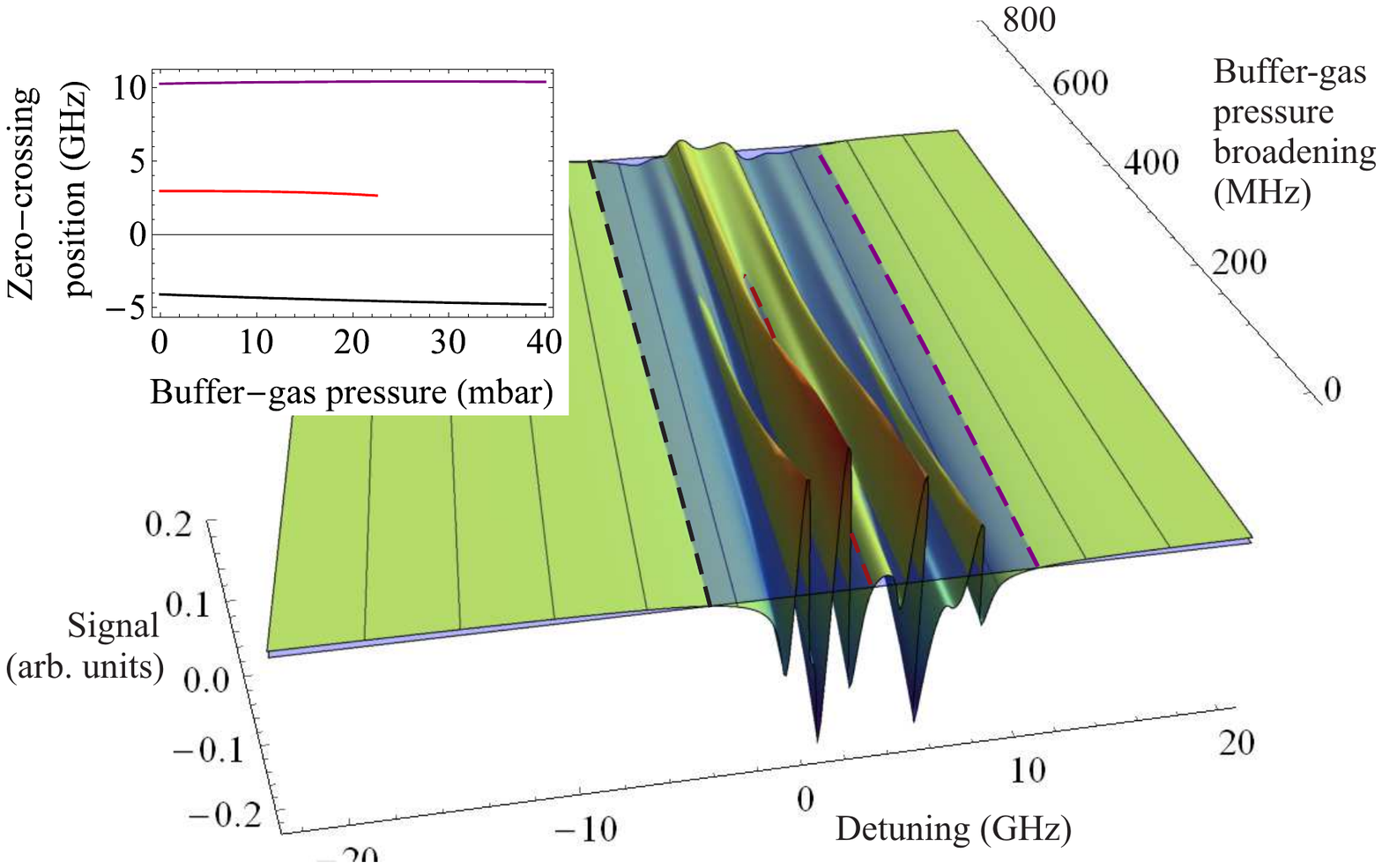}
	\caption{DAVLL spectra simulated as a function of buffer-gas pressure. The plot demonstrates the modification of the DAVLL spectra, particularly, signal deterioration, reshaping (the shaded areas highlight the signal with negative value), and associated reduction of zero crossings, with the buffer-gas pressure. The dashed lines mark the positions of three arbitrarily selected crossing points. The changes in the positions of these zero crossings are shown in the inset. The inset also reveals the disappearance of one of the crossing (marked with red line), resulting from interference of neighboring pressure-broadened transitions. The signals were simulated for $\theta_{\lambda/2}=0^\circ$ and $\theta_{\lambda/4}=-22.6^\circ$ (small deviation from perfect birefringence arrangement), at a temperature of about 50$^\circ$C, a magnetic field of 200~G, and a cell length of 1~mm.\label{fig:ConcentrationDependenceFull}}
\end{figure}
In such a case, the most pronounced effect of the buffer gas is signal deterioration and reshaping. Particularly, over the simulated range of buffer-gas pressures, the signal amplitude is reduced by roughly an order of magnitude. The reshaping can be demonstrated by the analysis of the zero crossings of the DAVLL signal (the shaded areas in the 3D plot mark regions with negative DAVLL signal). For example, the zero crossing observed at roughly 3~GHz (marked with red line) disappears at a N$_2$ pressure of about 20~mbar (see inset). The analysis also confirms the shift of the crossings. This is best visible at a low-frequency crossing (marked with black line), where the line broadening and shift cooperate toward shifting the resonance position further in low frequencies. At the same time, the shift of the high-frequency zero crossing (marked with purple lines) is much smaller as for such a tuning the two effects cause shifts in opposite directions. All these observations are in agreement with experimental data.

\begin{figure}
    \includegraphics[width=\columnwidth]{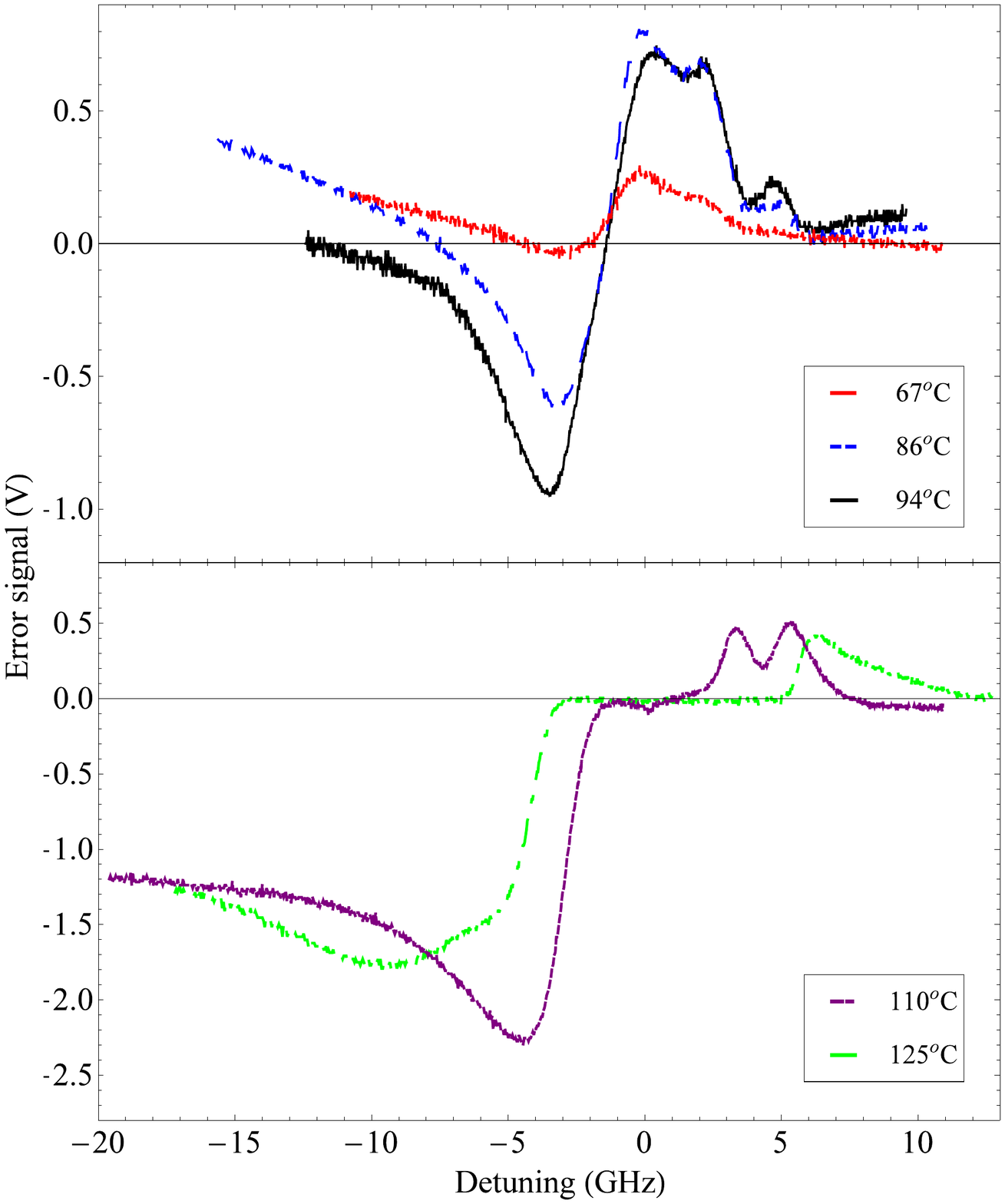}
    \caption{DAVLL signals measured at various vapor temperatures as a function of light-frequency detuning. At higher temperatures, the observed signal broadens due to stronger absorption of the signal (the change in the Doppler broadening is small, from 320~MHz at 67$^\circ$C to 340~MHz at 125$^\circ$C, so it may be neglected). Along with the broadening, the absorption in the cell also increases eventually leading to complete opacity of the vapor for light tuned close to the center of the transition (see, for examples spectra detected at 125$^\circ$C); under such conditions, laser cannot be stabilized for that frequency range.\label{fig:ErrorSignalOnTemperature}}
\end{figure}

A smaller number of zero crossings in the DAVLL spectrum of the buffer-gas-filled microcells with respect to the bulk cell may suggest limited locking range of the microDAVLL system with respect to the conventional systems. To demonstrate that this is not the case, we first investigate the temperature dependence of the 230-mbar microDAVLL spectra (Fig.~\ref{fig:ErrorSignalOnTemperature}).
As shown in in the figure, at the lowest temperature (67$^\circ$C), the DAVLL signal provided continuous stabilization within a $\sim$15-GHz range (detuning roughly from -10~GHz to 5 GHz), which was achieved by the combination of the Doppler ($\sim\! 0.3$~GHz) and pressure ($\sim\! 2.8$~GHz) broadenings. Locking at the particular tuning was obtained by a proper adjustment of wave plates. Particularly, locking the laser close to the maximum or minimum of the dichroic signal (i.e., at 0~GHz or -4~GHz in the considered case), where shown signals were flat, was possible by switching to detecting birefringence (rotating the quarter-wave plate by $\sim 45^\circ$) \cite{Yashchuk2000Laser,Marchant2011Off,Lee2011Small}. In fact, at such an arrangement, the birefringence provides the steepest error signals for these detunings, offering the strongest frequency locking.

Despite the ability to lock the laser frequency in a whole 15-GHz range using a 67$^\circ$C vapor, a problem of the system may be relatively small signal-to-noise ratio (SNR), which may result in weaker locking and hence more significant drifts of laser frequency) fluctuation of laser frequency. To avoid this problem, operation at higher vapor temperatures (larger optical depths) is desired. Importantly, at such temperatures, not only the signal becomes stronger but also the locking range increases. For instance, at $94^\circ$C, one can tune the stabilization point of the laser within a $\sim 20$-GHz range (a 5-GHz increase of the locking range). Further increase of the temperature (>100$^\circ$C) resulted on the one hand in extension of the wings of the DAVLL signal, on the other, however, in stronger absorption of the vapor. In particular, the signals measured at 110$^\circ$C and 125$^\circ$C reveal the loss of the ability to lock the laser close to the center of the $D_1$ line, due to the complete opacity of the vapor for such a frequency range.

To investigate the temperature dependence more thoroughly, we simulated the DAVLL spectra versus temperature for two different vapor cells: the evacuated cell and the cell filled with 1000~mbar of nitrogen. 
\begin{figure}
x	\includegraphics[width=9cm]{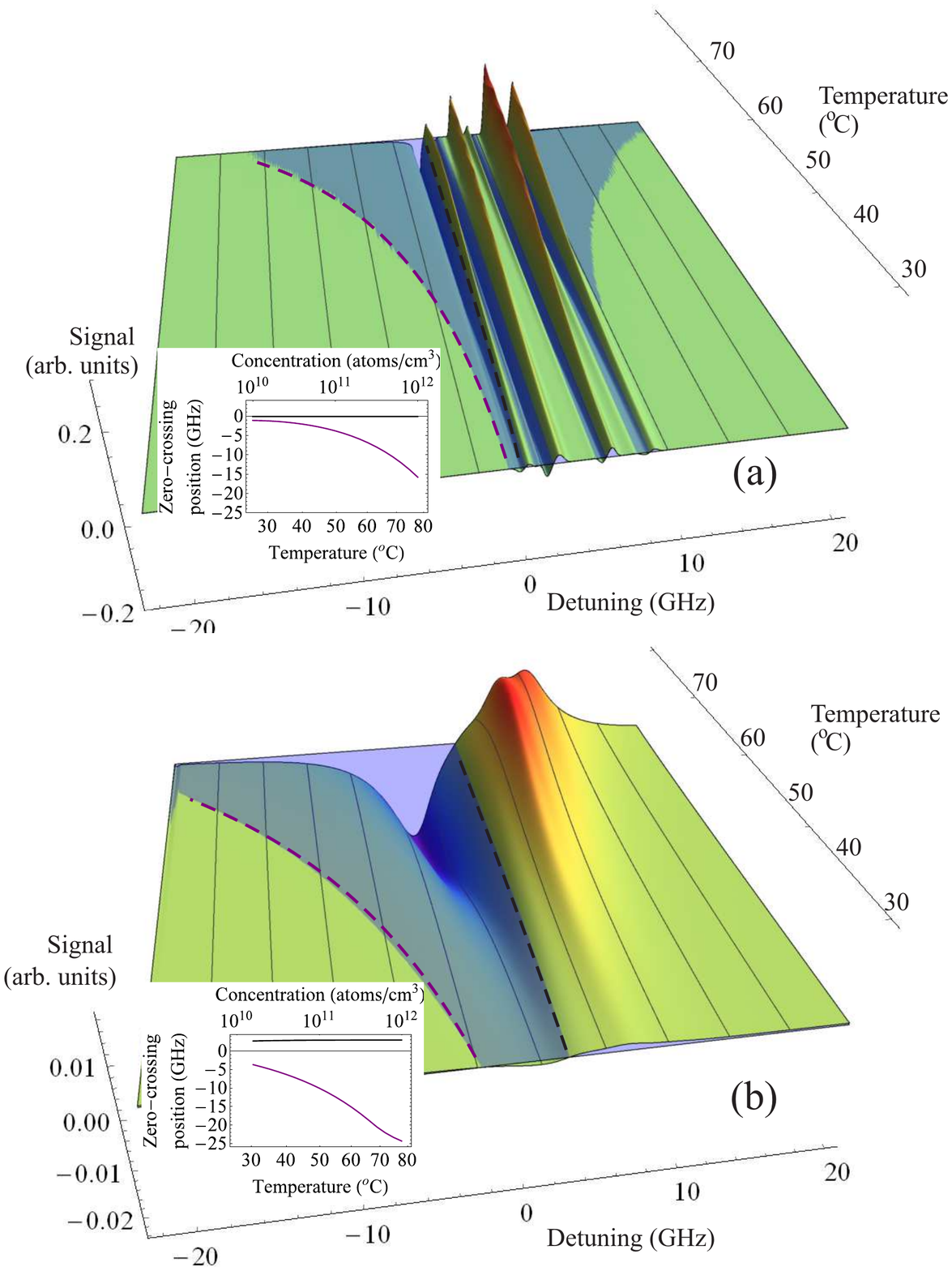}
	\caption{The DAVLL spectra measured versus temperature for the evacuted cell (a) and the cell filled with 1000~mbar of nitrogen (b). Increasing the temperature up to 70$^\circ$C enables enhancing the amplitude of the spectra. Above that temperature, the central part of the spectra starts to deteriorate, which is a result of absorption of light in optically thick medium. The temperature increase also leads to the extension of the high- and low-frequency wings of the signal, enabling shifting the locking points (see insets). The shaded areas mark regions of negative DAVLL signal. The simulations were performed in the dichroic geometry with $\theta_{\lambda/2}=45.3^\circ$ and $\theta_{\lambda/4}=45.8^\circ$ and a magnetic field of 200~G.\label{fig:ConcentrationDependenceFull}}
\end{figure}
The results of these simulations, shown in Fig.~\ref{fig:ConcentrationDependenceFull}, reveal general trends in the DAVLL-signal-temperature dependence, but also some differences between the two cases. Particularly, the simulations demonstrate that rising temperature, i.e., increasing atomic concentration, results in the enhancement of the DAVLL signal and extension of its wings toward high or low frequencies, independently from the cell type. For example, in the simulated temperature range (between 30$^\circ$C and 80$^\circ$C), the lowest frequency zero crossing was pushed, in both cells, more than 20~GHz to the lower frequencies (see insets). For even higher temperatures, the shifted is accompanied by oscillation of the low (and high) frequency part of the spectrum (not shown). This oscillations are a result of larger than 45$^\circ$ rotation of light polarization and were previously studied in Ref.~\cite{Marchant2011Off}. It is noteworthy, however, that the temperature induced increase of signal amplitude is limited. Particularly, in the central part of the spectrum, the DAVLL signal levels up at appropriately high temperatures (here at $\approx 70^\circ$C) and deteriorates for even higher temperatures (just a beginning of this trend is visible in Fig.~\ref{fig:ConcentrationDependenceFull}). This behavior stems from the absorption of the light in the medium and is responsible in the evacuated cell [Fig.~\ref{fig:ConcentrationDependenceFull}(a)] for lack of the ability to lock the laser in the central part of the spectrum, i.e., the frequency range between $F=3\rightarrow F'$ and $F=2\rightarrow F'$ transitions. However, access to this frequency range is possible with the cell filled with a buffer gas, which induces appropriately high pressure broadening [see Fig.~\ref{fig:ConcentrationDependenceFull}(b)]. In such a case, the $F=3\rightarrow F'$ and $F=2\rightarrow F'$ transitions overlap even at room temperature, so large increase in temperature is not required to observe coverage of rotation from the transitions (although may be desired to increase signal-to-noise ratio). 

In addition to the temperature dependent shifts of zero crossings, observed particularly at the wings of the spectra, the simulations also show that some of the crossings are unaffected by temperature (see, for example, the zero crossing observed at 1~GHz, marked in black in Fig.~\ref{fig:ConcentrationDependenceFull}). In that case, the crossings stay nearly unaffected, at positions of absorption lines unperturbed by the magnetic field ($B=0$).

\begin{figure}
    \includegraphics[width=\columnwidth]{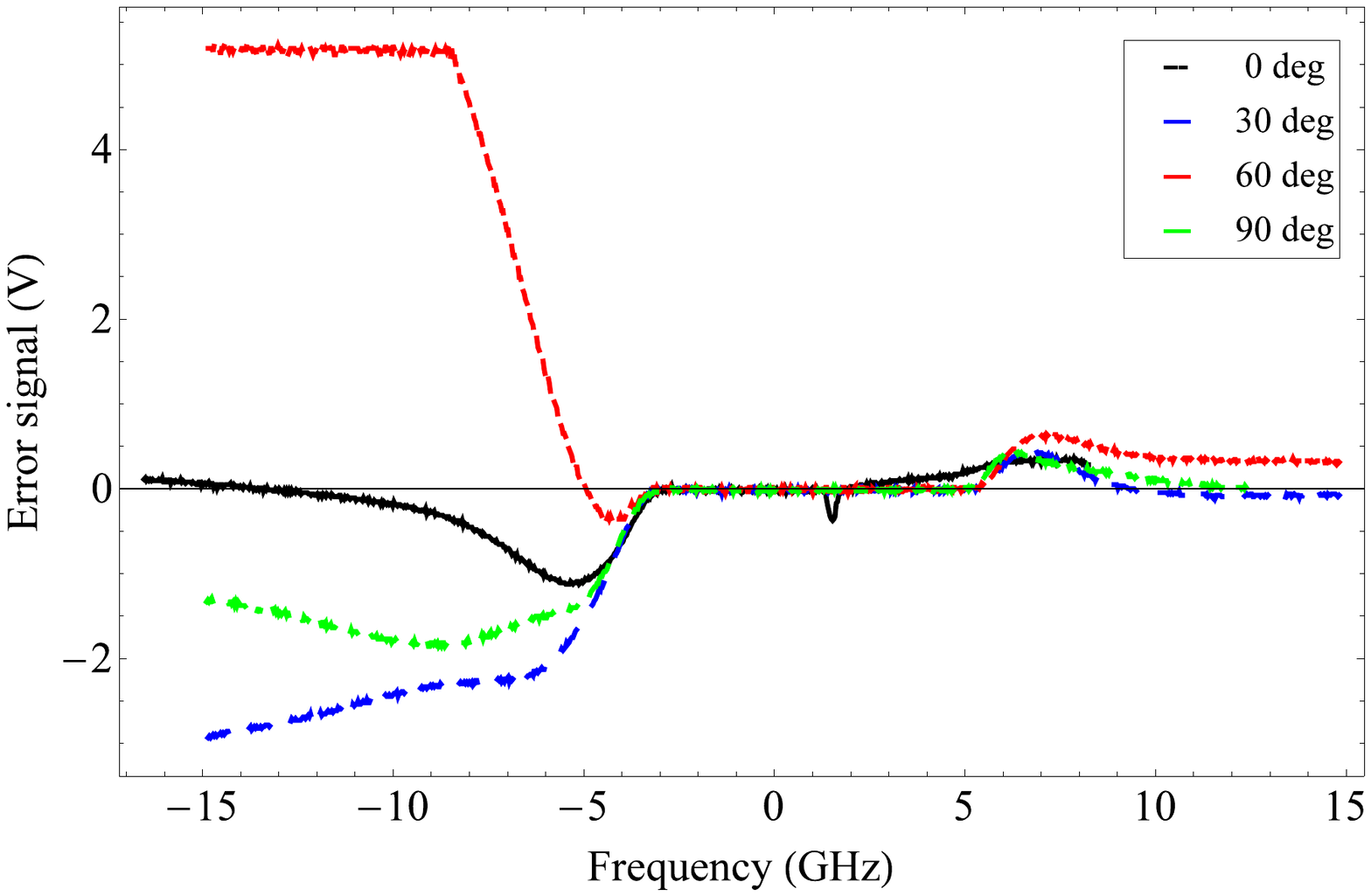}
    \caption{DAVLL spectra measured for different orientation of the quarter- and half-wave plates. Rotation of the wave plates modifies the DAVLL signal and the laser can be precisely locked anywhere between -15~GHz and -4~GHz and 8~GHz and 13~GHz. The constant signal of 5~V appearing in the 60$^\circ$ data at the detunings smaller than -8~GHz is due to the saturation of the detectors in our system. The signals were measured at the 230-mbar cell at a temperature of 125$^\circ$C. \label{fig:ErrorSpectrumAngle}}
\end{figure}

To further investigate the capability of the DAVLL system to lock the laser far from the optical transition, we also studied the dependence of the signal on the orientation of quarter-wave plate. Figure~\ref{fig:ErrorSpectrumAngle} shows the error signal measured as a function of the light frequency for various quarter-wave-plate orientations\footnote{Every quarter-wave-plate rotation results in reshaping of the signal, particularly, shift of its base line. Thus, to enable locking of the laser, i.e., to observe zero crossings of the signal, every rotation of the quarter-wave plate is accompanied by an appropriate rotation of the half-wave plate.}.
The signals were measured in the 230-mbar cell at a temperature of 125$^\circ$C, where no transmission is observed at the center of the $D_1$ line. At the same time, the wings of the signal extend far into lower and higher frequencies ($\gtrsim$10~GHz) enabling off-resonance locking of the laser. In particular, for the $0^\circ$ orientation of the wave plate (relative angle), the locking point (the zero crossing) appears $\sim 15$ GHz away from the closest optical transition ($F=2\rightarrow F'=1$ of $^{87}$Rb). The position of this point may be precisely adjusted by rotating the quarter-wave plate so that the laser can be stabilized to any frequency within the low-frequency part of the spectrum. Specifically, at $60^\circ$ the locking point can be brought 10 GHz closer to the center of the line (being still ${-4}$ GHz below the center of the $F=2\rightarrow F'=1$ transition of $^{87}$Rb). Similarly, one can lock the laser $\sim$3~GHz above the $^{87}$Rb $F=1\rightarrow F'=2$ transition (30$^\circ$) and the tuning range extends roughly 5~GHz further into the higher frequencies for 90$^\circ$ quarter-wave-plate orientation.

It is noteworthy that the locking range at positive detunings is smaller than that obtained for negative detunings. This is in part due to the difference in the shift of the transition frequencies due to the buffer gas (pressure shift). As shown in Fig.~\ref{fig:ErrorSpectrumPressure}, the transitions in the 230-mbar cell are shifted by --1.3~GHz with respect to the evacuated cell. Hence the whole locking range is 1.3-GHz detuned toward the lower frequencies. Additionally, the combined strength of the $^{87}$Rb $F=2\rightarrow F'$ and $^{85}$Rb $F=3\rightarrow F'$ transitions is larger than that of the $^{85}$Rb $F=2\rightarrow F'$ and $^{87}$Rb $F=1\rightarrow F'$ transitions, which are weaker and further apart. This leads to the larger amplitude of the DAVLL signal at the low-frequency wing compared to that at the high-frequency wing of the spectrum and thus lower range of frequencies for laser stabilization. Finally, there is a contribution to low-high-frequency asymmetry from the change of the light intensity accompanying the sweep of the light frequency (achieved by laser-current modulation). In turn, the amplitude of the low-frequency signal is larger than that of high frequency. Over the frequency-sweep range of Fig.~\ref{fig:ErrorSpectrumAngle}, the change is on the order of 20\%. The slope associated with this effect is most pronounced in the 30$^\circ$ data.

\begin{figure}
	\includegraphics[width=9cm]{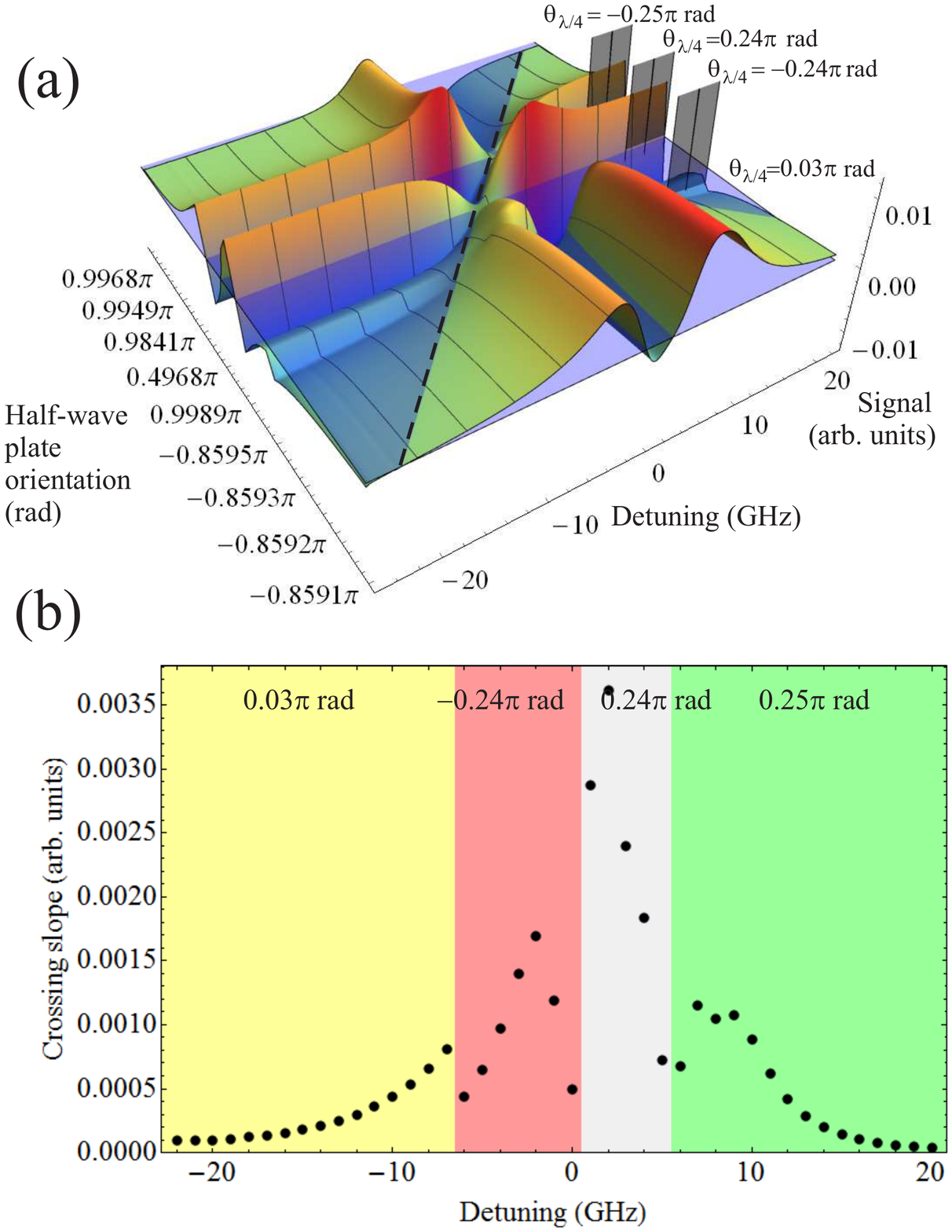}
	\caption{DAVLL spectra for different wave-plate orientations (a) and slope of the spectra for different tunings and wave-plate orientations (b). The black polygons at the right-hand side of the 3D plots separate regions with different quarter-wave-plate orientations. Within the region the half-wave plate orientation is changed smoothly between two extreme values. The dashed line marks the position of zero crossings and the shaded area highlights the area with negative DAVLL signal. Rotating the wave plates enables controllable tuning of the laser lock point (zero crossing) in a continouos 40~GHz range. While tuning the locking point, the steepness of the slope changes nearly by 2 orders of magnitude, changing the strength of lock, i.e., system's ability to neutralize frequency drifts or instant jumps (the label in the plot denotes the quarter-wave plate orientation). This loss can be compensated with the increase of vapor temperature or gain in the electronic system. The simulations are performed for temperature of 50$^\circ$C, nitrogen pressure of 1000~mbar, cell length of 1~mm, and magnetic field of 200~G. \label{fig:WaveplateDependenceFull}}
\end{figure}

To fully demonstrate the capabilities of the DAVLL system exploiting the buffer-gas cells, we investigated a position of zero crossing on the orientations of the wave plates [Fig.~\ref{fig:WaveplateDependenceFull}(a)].
With the 1000-mbar cell, we demonstrated the ability to continuously tune the lock point of a laser in a 40-GHz frequency range. Similarly as in a traditional DAVLL system, this range is not achieved with a single orientation of wave plates, but requires their appropriate adjustments. Particularly, in the presented case, four orientations of the quarter-wave plate and many (principally, infinite number of) orientations of a half-wave plate were used. The changes of the quarter-wave-plate orientation were applied to switch between Faraday and dichroic geometries, as well as, reverse the sign of the signal. Rotation of the half-wave plate is used to shift the zero-crossing point. In turn, the half-wave-plate rotation induces small modification of the spectra, while rotation of the quarter-wave plate their significant change (note different region marked by black polygons in the plot). 

An interesting question of locking a laser far from an optical transition is the strength of such lock. This strength is determined by the steepness of the DAVLL signal. To investigate this, we calculated the slope of the DAVLL signal in the 1000-mbar cell for the zero crossings separated by 1~GHz [Fig.~\ref{fig:WaveplateDependenceFull}(b)]. As shown, the slope steepness changes roughly 30 times between the strongest locking (close to the center of the pressure-broadened absorption line, i.e., at a 0-GHz detuning) and the weakest locking points (at far low/high-frequency wings of the transition, i.e., for a $\pm$20-GHz detunings). As shown above, the steepness can be enhanced either by increasing the temperature of the vapor or by improving electronic gain. Each of this solution has its advantages and disadvantages. Particularly increasing temperature may increase a signal at the wings but reduce its amplitude at the center (due to enhanced absorption) and improving electronic gain is simple but keeps signal-to-noise ratio at the same level, so the noise, and associated laser-frequency fluctuations, can be significant. Therefore, a careful choice of experimental parameters is required for correct operation of the system under given conditions.

\begin{figure}
    \includegraphics[width=\columnwidth]{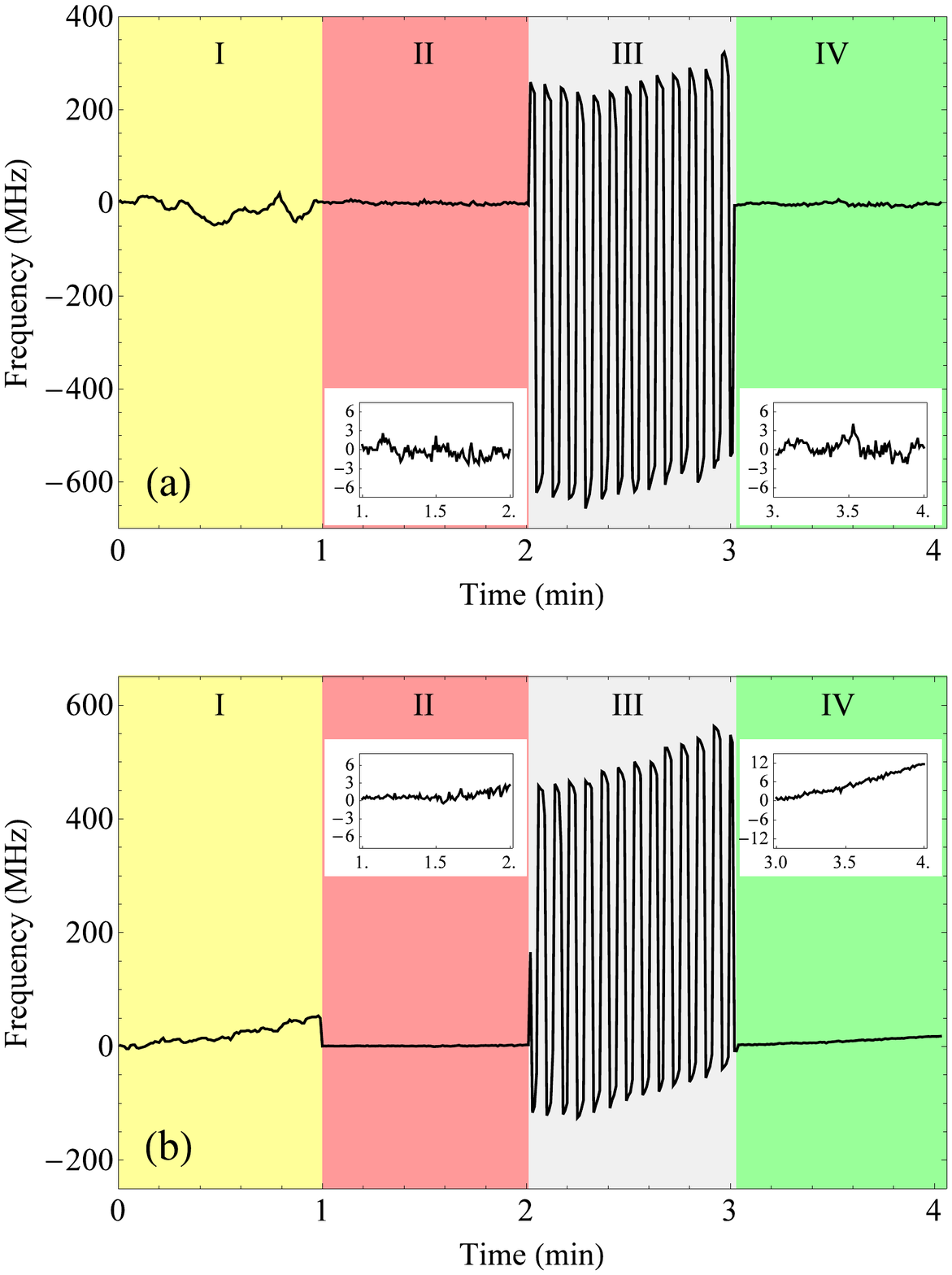}
    \caption{The performance of the DAVLL system for the laser locked between the $F=3\rightarrow F'$ and $F=2\rightarrow F'$ transitions of $^{85}$Rb (a) and 16~GHz away from the unperturbed $F=2\rightarrow F'=1$ transition of $^{87}$Rb (b). The laser stabilization was performed using the 230-mbar-cell DAVLL system at 110$^\circ$C. From 0 to 1 min (yellow, I), the laser frequency was not stabilized and freely drifted over time. Between minute 1 and 2 (red, II) the laser frequency was locked using the DAVLL system. Over the next minute (blue, III), the laser was unlocked and modulation was put into the modulation channel of the laser. The current modulation corresponded to 600~MHz (peak-to-peak). Over the last minute (green, IV), the DAVLL stabilization was turned on with the applied current modulation. In such a way, the ability of the locking system to compensate for instant up to 600-MHz frequency jumps (peak-to-peak) was demonstrated. The inset showing the magnification of the stabilization signal reveals a stability better then 5~MHz. The positive slope observed toward the end of the 4$^\textrm{th}$ minute is due to the temperature-induced shift of the zero-crossing position (see text).\label{fig:Stabilization}}
\end{figure}

To experimentally demonstrate the ability to lock the laser at frequencies unattainable for conventional DAVLL systems, we used a microDAVLL system exploiting the 230-mbar vapor cell (Fig.~\ref{fig:Stabilization}).
The figure depicts two cases of laser locking: between the $F=3\rightarrow F'$ and $F=2\rightarrow F'$ transitions of $^{85}$Rb [Fig.~\ref{fig:Stabilization}(a)] and 16~GHz away from the $F=2\rightarrow F'=1$ transition of $^{87}$Rb [Fig.~\ref{fig:Stabilization}(b)]. In both cases four regimes are shown: laser unlocked (I), laser locked (II), laser unlocked with superimposed modulation (III), and laser locked with superimposed modulation (IV). As shown in cases (II) and (IV), the DAVLL was able to compensate for slow frequency drifts of the signal as well as instant frequency changes/jumps (with a 600-MHz amplitude peak-to-peak). For both laser lock points, the system reveals the same stability (better then 5~MHz) for the time of an hour (not shown).

Analysis of the locking signal shown in Fig.~\ref{fig:Stabilization}(b) (particularly toward the end of the measurements) reveals a slow drift of the frequency of the locked laser. This drift occurred even though the error signal was kept at zero the whole time. Therefore, the drift does not result from lack of ability to compensate for the drift, but rather reflects shifting of the zero-crossing position of the DAVLL signal. To verify weather the drift was caused by temperature variation of the atomic vapor, we took the approach presented in Ref.~\cite{Kostinski2011Temperature}, where it was shown that an appropriate choice of the incident-light polarization makes the DAVLL system insensitive to the temperature changes. Despite a number of attempts, we were unable to demonstrate such insensitivity. In fact, the frequency drifts were comparable for all wave-plate arrangements. This led us to the conclusion that the drift arises not due to the change in atomic-vapor density but due to temperature-induced changes of the magnetic field the vapor is subject to or changes of the birefringence of the glass windows of the cell. We verified by numerical simulations that variation of the magnetic field by 10\% results in a shift of the locking point by not more than 300~MHz. The drift, however, strongly depends on the locking point, i.e., the further toward the wing the laser is locked, the more sensitive it is to magnetic-field drifts. These results show that for more reliable performance of the system under unusual conditions, e.g., for locking the laser far from the optical transition, a DAVLL with more stable magnetic field is required. A possible solution maybe a system where permanent magnets are replaced a magnetic-field coil. This is particularly appealing as the coil may also be used for heating the cell similarly as proposed in Ref.~\cite{Marchant2011Off}.

\section{Conclusion}

We discussed a new version of a DAVLL system that exploits buffer-gas-filled millimeter-scale vapor cells. The system offers similar stability as is achievable with bulk vapor cells, additionally offering several advantages. In addition to its compactness, the microDAVLL system with buffer-gas cells may provide continuous stabilization in a multi-gigahertz range around the optical transition. The range may be controlled either by changing the temperature of the vapor or by application of the buffer gas under an appropriate pressure. In particular, we demonstrated that the buffer gas enables locking the laser between the two hyperfine components of the $^{85}$Rb ground state or far from the center of optical resonance (-16~GHz from the center of $^{87}$Rb $F=2\rightarrow F'=1$ transition) with a residual instability of less than 5~MHz. The model developed for the system enabled calculation of the DAVLL signal and optimization of its operation (by adjustment of gas mixture, vapor temperature, and buffer-gas pressure) for specific applications. The ability to stabilize laser light far from optical transition will be useful in applications where light works as a weakly perturbing probe. A particular example of such application is nuclear magnetic resonance in zero or ultra-low magnetic fields \cite{Ledbetter2009Optical,Ledbetter2011Near}, where the probe laser needs to be detuned tens of GHz from the optical transition.

\begin{acknowledgments}
We thank Stefan Woetzel for the manufacturing of the Rb cells and Simon Rochester for stimulating discussion and suggestions regarding simulations of the DAVLL signal using the AtomicDensityMatrix Mathematica package. This work was supported by the National Centre for Research and Development within the Leader Program and by the National Science Foundation under Award CHE-1308381.
\end{acknowledgments}

\end{document}